\documentclass[aps,prl,twocolumn,superscriptaddress,showpacs,showkeys]{revtex4}

\usepackage{graphicx}

\usepackage{dcolumn}

\usepackage{bm}

\begin{document}








\title{Manipulation of electrical and ferromagnetic properties of photo-sensitized (Ga,Mn)As\\}






\author{L. Herrera Diez}

\email{l.herrera-diez@fkf.mpg.de}

\affiliation{Max-Planck-Institut f\"ur
Festk\"orperforschung, Heisenbergstrasse 1, 70569, Stuttgart, Germany}

\author{M. Konuma}
\affiliation{Max-Planck-Institut f\"ur
Festk\"orperforschung, Heisenbergstrasse 1, 70569, Stuttgart, Germany}

\author{E. Placidi}%
\affiliation{Dipartimento di Fisica, Universit\`a di Roma "Tor
Vergata", and CNR-INFM, Via della Ricerca Scientifica 1, I-00133
Roma, Italy}

\author{F. Arciprete}%
\affiliation{Dipartimento di Fisica, Universit\`a di Roma "Tor
Vergata", and CNR-INFM, Via della Ricerca Scientifica 1, I-00133
Roma, Italy}

\author{A. W. Rushforth}%
\affiliation{School of Physics and Astronomy, University of
Nottingham, University Park, Nottingham NG7 2RD, United Kingdom}

\author{R. P. Campion}%
\affiliation{School of Physics and Astronomy, University of
Nottingham, University Park, Nottingham NG7 2RD, United Kingdom}

\author{B. L. Gallagher}%
\affiliation{School of Physics and Astronomy, University of
Nottingham, University Park, Nottingham NG7 2RD, United Kingdom}

\author{J. Honolka}%
\affiliation{Max-Planck-Institut f\"ur
Festk\"orperforschung, Heisenbergstrasse 1, 70569, Stuttgart, Germany}

\author{K. Kern}%
\affiliation{Max-Planck-Institut f\"ur
Festk\"orperforschung, Heisenbergstrasse 1, 70569, Stuttgart, Germany}
\affiliation{Institut de Physique des Nanostructures,
Ecole Polytechnique F\'ed\'erale de Lausanne,\\ CH-1015 Lausanne, Switzerland.}
\date{\today}

\begin{abstract}

We present the manipulation of magnetic and electrical properties of
(Ga,Mn)As by the adsorption of dye-molecules as a first step towards
the realization of light-controlled magnetic-semiconductor/dye
hybrid devices. A significant lowering of the Curie temperature with
a corresponding increase in electrical resistance and a higher
coercive field is found for the GaMnAs/fluorescein system with
respect to (Ga,Mn)As. Upon exposure to visible light a shift in
Curie temperature towards higher values and a reduction of the
coercive field can be achieved in photo-sensitized (Ga,Mn)As. A
mayor change in the XPS spectrum of (Ga,Mn)As indicates the
appearance of occupied levels in the energy range corresponding to
the (Ga,Mn)As valence band states upon adsorption of fluorescein.
This points towards a hole quenching effect at the
molecule-(Ga,Mn)As interface which is susceptible to light exposure.

\end{abstract}






\maketitle The discovery of the ferromagnetic semiconductor
(Ga,Mn)As~\cite{ohno} with remarkable properties such as hole
mediated ferromagnetism~\cite{dietl} has motivated an intense
research activity in view of potential applications in spintronics.
The material's most striking characteristic is the manifold
possibilities to tune magnetic properties via the manipulation of
the carrier density or the modulation of the lattice strain. A
non-volatile modulation of the carrier density can be induced by
means of ferroelectric gating~\cite{ferroelec}. The use of
conventional metallic gates in a classic field effect transistor
geometry on the other hand allows for the continous tuning of the
magnetization in the absence of magnetic fields~\cite{gating}.
Strain controlled ferromagnetism in (Ga,Mn)As has been equally
exploited to manipulate magnetic anisotropy in systems showing
lithography-induced anisotropic strain
relaxation~\cite{molenkamp,wunderlich} and in GaMnAs/piezoelectric
actuator hybrid structures~\cite{strainandrew,groennenwein}. In this
study we present the effect of light-sensitive molecular layers
adsorbed on thin (Ga,Mn)As epilayers. We use fluorescein and two of
its derivatives to achieve an effective hole quenching in the
magnetic semiconductor and show, as a proof of principle, the
possibility to influence this hole quenching process by light
excitation. Our result on GaMnAs/organic dye hybrid systems is an
important first step towards magnetic
applications exploiting molecule-dependent, photon-controlled carrier modulation.\\
The (Ga,Mn)As films used in this study are grown by molecular beam
epitaxy (MBE) on (Ga,Mn)As substrates in two different MBE
laboratories. The two sample materials designated A and B with
thicknesses of 50nm and 40nm exhibit Curie temperatures $T_c$ of 68K
and 48K, respectively. The Mn concentrations for samples A and B are
5\% and 8\%, respectively. Further details regarding the epitaxial
growth can be
found in Ref.[7] and [8] for sample A and B, respectively.\\
The molecular layers were adsorbed by immersing the (Ga,Mn)As films
in a 2~mM solution of the molecules. This solution is prepared by
dissolving the molecule powder in water at room temperature and then
adjusting the pH to 7 with NaOH. Prior to the immersion in the
solution the (Ga,Mn)As films are treated with an HF containing etch
mixture (Original etch mixture: AF 87.5-12.5 VLSI Selectipur,
diluted 1:100 in water) for approximately 10 seconds to clean the
surface. After HF treatment the samples are rinsed in purified water
and thereafter immediately placed in the molecule solution for about
12 hours. Finally, the samples are rinsed in water to remove most of
the non-chemisorbed species and are left to dry in air at room
temperature and protected from visible light. Magneto-transport
measurements performed before and after the HF treatment confirm
that the etching procedure does not modify the properties of the
(Ga,Mn)As films. The light source employed in the illumination
experiments is a standard HBO Hg lamp. No light-induced effects or
heating have been observed in the magneto-transport
properties in as grown (Ga,Mn)As films without molecules.\\
The expected adsorption geometry of fluorescein and its chemical
structure is illustrated in the inset of Fig. 1 (b) and in Fig. 2
(top), respectively. The molecule is thought to interact with the
(Ga,Mn)As surface mainly via the carboxylic group as proposed in the
literature for the chemisorption of carboxyl-terminated molecules on
GaAs/GaMnAs heterostructures~\cite{molecule} and for the adsorption
of fluorescein and other aromatic acids on semi-insulating and
metallic surfaces~\cite{fluorescadsorbed}.\\
The X-ray photoelectron spectra (XPS) of sample A before and after
fluorescein adsorption are shown in Fig. 1 (a) and (b),
respectively. In the spectrum of the as-grown sample the dominating
feature is the Ga 3d peak (19.6 eV) and a shoulder towards higher
binding energies (21.3 eV) accounting for the presence of gallium
oxide \cite{galliumoxide,nota}.

\begin{figure}
\includegraphics[width=8cm]{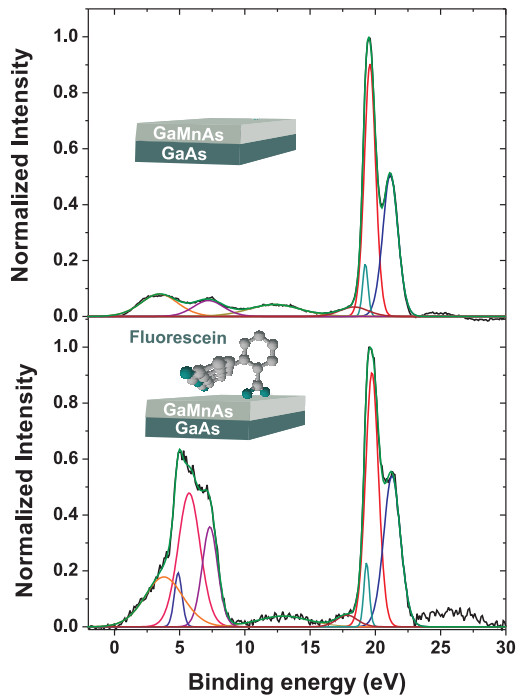}\\
\vspace{-0.6cm}
\caption{Normalized XPS spectra of the Ga 3d peak and the (Ga,Mn)As
valence band before (a) and after (b) adsorption of fluorescein. A
distinct feature composed of several peaks (see text) centered
around 5~eV appears in spectrum (b) indicating the appearance of occupied electronic levels in the energy
range of the (Ga,Mn)As valence band states upon adsorption of the
molecule. (Inset) Sketch of the tentative adsorption geometry of
fluorescein.}\label{one}
\vspace{-0.2cm}
\end{figure}
The mayor changes in the spectra related to the adsorption of
fluorescein occur at lower binding energies with respect to the Ga
3d peak, corresponding to the valence band states in bulk
GaAs(001)\cite{valenceband} and by analogy also to the valence band
of its p-doped variant (Ga,Mn)As. A distinct increase in the
intensity of the existing peaks (around 7.3eV and 3.4 eV) relative
to the Ga 3d peak and the appearance of new features (at 5.7 and 4.9
eV) indicate the appearance of occupied electronic levels in the
energy range of the (Ga,Mn)As valence band states upon adsorption of
fluorescein. The new lineshape possibly reflects the changes
occurring in the GaMnAs states due to the interaction with the
adsorbates together with the occupied states of the molecule. In any
case, the appearance of populated levels at these relatively low
binding energies could point at an effective hole quenching in the
(Ga,Mn)As substrate that according to the theory of hole mediated
ferromagnetism\cite{dietl} would lead to
the weakening of the ferromagnetic interaction.\\
\begin{figure}
\includegraphics[width=8cm]{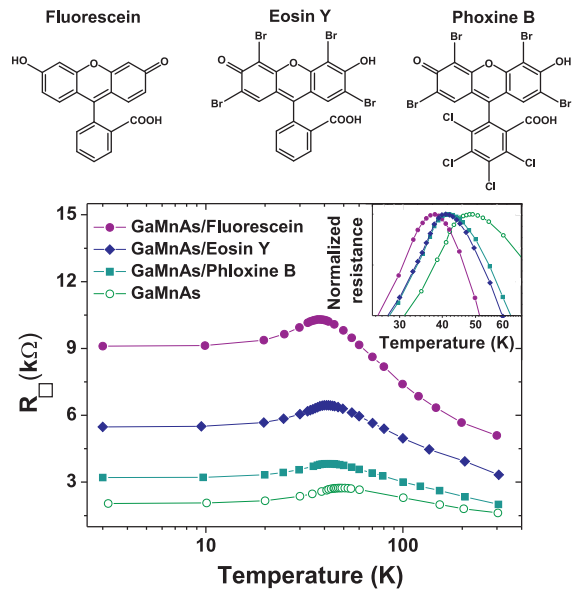}\\
\vspace{-0.6cm}
\caption{Effect of fluorescein and its derivatives eosin Y and
phloxine B on the magnetic and electrical properties of sample B.
The introduction of electronegative groups in the structure of the
molecules reduces the hole quenching capability resulting in a smaller shift in $T_{c}$ and a
reduced increase in electrical resistance with respect to
fluorescein.}
\label{one}
\vspace{-0.2cm}
\end{figure}
Presented in Fig. 2 is the temperature dependence of the electrical
resistance showing the characteristic peak (inset) indicative of the
Curie temperature\cite{peakresistance} for the interaction between
(Ga,Mn)As (sample B) and fluorescein derivatives. In line with the
picture of a hole compensation mechanism derived from the XPS
spectra an increase in the electrical resistance and a shift of the
resistance peak towards lower temperatures is observed for all the
molecules, fluorescein producing the largest effect. The two
fluorescein derivatives investigated are eosin Y (diamonds) and
phloxine B (squares) that differ from fluorescein by additional
electronegative bromide and chlorine groups as indicated in the
molecular structure shown in Fig 2. The presence of electronegative
groups is expected to hinder the hole quenching capability of the
molecule and is in agreement with our experiments. The adsorption of
eosin Y produces a smaller decrease in the Curie temperature and a
reduced increase in the electrical resistance with respect to
fluorescein. An even less efficient hole quenching is observed for
phloxine B where a slightly higher Curie temperature and a smaller
resistance
increase is observed with respect to eosin Y.\\
In Fig. 3 the coercive fields \emph{H$_{c1}$} (circles) and
\emph{H$_{c2}$} (squares) at a temperature of 1.5 Kelvin for
(Ga,Mn)As (sample B, full symbols) and GaMnAs/fluorescein (open
symbols) as a function of the direction of the applied magnetic
field with respect to the [110] uniaxial easy axis direction are
displayed. The magnetic anisotropy landscape given by the coercive
fields map shows a rectangular shape with its long axis parallel to
the [110] direction for \emph{H$_{c1}$} and \emph{H$_{c2}$}
approaching the [110] and $[1\bar{1}0]$ axes which is typical for
as-grown (Ga,Mn)As materials~\cite{tang,nuestro}. The origin of this
two step reversal is the two fold symmetry in the magnetic
anisotropy of GaMnAs/GaAs given by the interplay between a biaxial
magneto-crystalline anisotropy (easy axes along [100] and [010]
directions) and a
uniaxial anisotropy (easy axis along [110]) component.\\
\begin{figure}
\includegraphics[width=5cm]{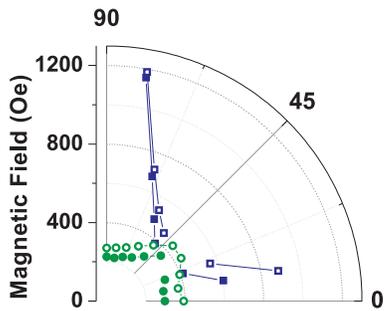}\\
\caption{Coercive fields \emph{H$_{c1}$}(circles) and
\emph{H$_{c2}$}(squares)
as a function of the angle of the applied magnetic field with
respect to the [110] direction at 1.5 K for sample B. An overall
increase in the coercivity in all directions is observed upon
adsorption of fluorescein. The coercivities of (Ga,Mn)As and GaMnAs/Fluorescein
are indicated with full and open symbols, respectively. }\label{one}
\vspace{-0.3cm}
\end{figure}
After adsorption of the molecules an overall increase of the
coercive fields in all directions can be observed in agreement with
the lower value of $T_{c}$ as it has been reported by other
authors~\cite{coercivity}. A change in the magnetic anisotropy upon
fluorescein adsorption is evidenced in a modest strengthening of the
uniaxial easy axis ([110] axis, 0$^{\circ}$ direction) where the
increase in coercivity is larger than along the uniaxial hard axis
direction ($[1\bar{1}0]$ axis, 90$^{\circ}$ direction). The ratio
$H_{c}$$[1\bar{1}0]$/$H_{c}$[110] decreases from 0.75 for the as
grown (Ga,Mn)As to 0.71 for GaMnAs/fluorescein indicating an
approximately 5$\%$ increase in the easy axis character of the [110]
direction. It has been shown both theoretically and experimentally
that not only the Curie temperature but also the magnetic anisotropy
in (Ga,Mn)As is susceptible to variations of the carrier
density~\cite{dietl,gating,gallagherdos}. However, theory and
experiments disagree when determining the anisotropy component
(uniaxial or biaxial) that is most affected by changes in carrier
concentration. Experimental work shows that the uniaxial term is
most sensitive to variations in the carrier concentration, which can
lead to a 90$^{\circ}$ change in the orientation of the uniaxial
easy axis (from $[\bar{1}10]$ to [110]) when reaching a critical
value of the carrier density. A change of sign in the uniaxial
magnetic anisotropy energy is not observed in our study probably
because we are probing a hole concentration range that is far from
the critical value. Nevertheless, the observation of a change in the
uniaxial anisotropy component with respect to the biaxial component
upon carrier concentration is clearly in line with results in the
literature. The general increase of coercivities in all directions
can be discussed in terms of micromagnetic models. Since the
magnetization reversal in this material is known to occur by domain
wall motion~\cite{nuestro} a common expression for the domain wall
pinning energy density ($\epsilon$) can be included in this
analysis~\cite{tang}: $\epsilon=(\textbf{M}_{2}-
\textbf{M}_{1})\cdot \textbf{H}_{c}$, where $\textbf{M}_{1}$ and
$\textbf{M}_{2}$ are the initial and final magnetization vectors and
$\textbf{H}_{c}$ is the coercive field. Keeping in mind the small
changes in the magnetic anisotropy after the adsorption of
fluorescein we can assume that $\epsilon$ remains approximately
constant for every direction. Therefore, according to the above
expression a general increase in $\textbf{H}_{c}$ could be directly
related to a decrease in the absolute magnetization value $M$ that
is in agreement with the notion of a weakened ferromagnetic
interaction given by the smaller value of Curie temperature after
adsorption of fluorescein.\\
\begin{figure}
\includegraphics[width=9cm]{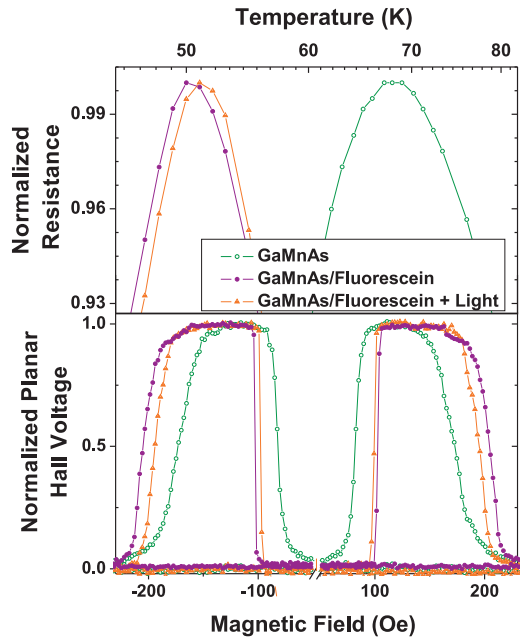}\\
\vspace{-0.6cm}
\caption{Light induced changes in the magnetic properties of
GaMnAs/fluorescein. A shift in the Curie temperature towards higher
values is observed upon illumination (top) together with a decrease
in the value of the coercive field (bottom).}\label{one}
\vspace{-0.2cm}
\end{figure}
The shift in $T_{c}$ upon adsorption of fluorescein for sample A is
18 Kelvin as shown in Fig. 4 and like in the case of sample B it is
accompanied by an increase in resistance. It is useful to observe
that in the present case these two quantities can be linked together
by the well know expression $ T_{c} \propto p^{1/3}$ provided by
mean field theory\cite{dietl}, where $p$ is the hole carrier
density. The ratio
$\frac{T_{c}^{\text{GaMnAs}}}{T_{c}^{\text{GaMnAs/Fluorescein}}}$$\sim$1.36
is in good agreement with the ratio of the cubic root of the inverse
resistances (assumed to be proportional to $p$) which has a value of
approximately 1.35. A similar relation is found for the
case of sample B.\\
In the following we show results which prove that the
(Ga,Mn)As-fluorescein interaction introduces an additional degree of
freedom for magnetic manipulation that can be controlled by light. A
direct interaction between light and (Ga,Mn)As affecting the
magnetic properties has been demonstrated by the injection of
optically generated spin-polarized holes into the (Ga,Mn)As material
causing changes in the magnetization~\cite{opticalspin}. In our
approach we do not intend to manipulate the magnetic properties by
direct interaction between (Ga,Mn)As and light but instead modulate
the hole quenching capability of the fluorescein molecules as a sort
of light-regulated gate.\\
Illumination experiments are carried out by shining light on the
magneto-transport devices during both the cool down of the sample
and the entire magnetic measurements. Fig. 4 shows the effect of the
illumination on sample A after the adsorption of fluorescein. A
distinct shift of the resistance peak towards higher temperatures
during illumination can be observed in Fig. 4 (top) which is
accompanied by a decrease in the value of the resistance (not
shown). In addition, a decrease in the value of the coercive field
is observed in the field dependence of the planar Hall voltage as
shown in Fig. 4 (bottom). Thus, the illumination effect is exactly
inverse to the pure adsorption effect and according to our
discussion we attribute the effect of light to an increase of hole
carriers in (Ga,Mn)As. The mechanism behind the illumination effect
is possibly connected to the position in energy of the molecule
ground and exited states with respect to the (Ga,Mn)As valence and
conduction band edges. The molecule ground state is expected to be
located close to the (Ga,Mn)As valence band in order to be able to
effectively interact with the holes in these levels. Upon light
absorption a fraction of these electrons that were interacting with
the valence band are now being promoted to the excited state of the
molecule, and are no longer able to contribute to the hole
compensation producing a small increase in the hole concentration
and an increase in the Curie temperature.\\
The light effect is reversible upon removal of the light source
although the return to the original state occurs via a slow
relaxation mechanism on a time scale of hours. The physical
mechanism behind this relaxation still remains an open question,
nevertheless, it is important to observe the reversibility of the
effect that points at the good photostability of the adsorbates in
view of potential practical applications.\\
In conclusion, the results presented here demonstrate the
effectiveness of molecular layers in quenching hole carriers in
(Ga,Mn)As changing its magnetic properties. Most importantly this
study shows how this interaction can be modified by light absorption
of the molecular layers which is a first step towards
light-controlled ferromagnetism in GaMnAs/dye devices. The reported
observations provide a proof of principle for an effect that can be
further explored using electron-acceptor molecules or even enhanced
by functionalization with photoisomerizable molecules that can
additionally provide non-volatility.\\ \vspace{-0.2cm}

\subsection*{Acknowledgments}

We would like to thank A. Schulz for photoluminescence measurements
and K. Balasubramanian for valuable help with the etching procedure.
We also acknowledge useful discussion with M. Burghard and H. Klauk.

\end{document}